\documentclass[a4paper]{article}
\usepackage{amsmath,amssymb, amsthm}
\title{Von Neumann and Luders postulates and quantum information theory }
\author{Andrei Khrennikov\\
School of Mathematics and Systems Engineering\\
International Center of Mathematical Modeling\\ in Physics and Cognitive Sciences\\
V\"axj\"o University, S-35195, Sweden}

\begin{document}

\maketitle

\abstract{This note is devoted to some foundational aspects of
quantum mechanics (QM) related to quantum information (QI) theory,
especially quantum teleportation and ``one way quantum
computing.'' We emphasize the role of the projection postulate
(determining post-measurement states) in QI and the difference
between its L\"uders and von Neumann versions. These projection
postulates differ crucially in the case of observables with
degenerate spectra. Such observables play the fundamental role in
operations with entangled states: any measurement on one subsystem
is represented by an observable with degenerate spectrum in the
Hilbert space of a composite system. If von Neumann was right and
L\"uders was wrong the canonical schemes of quantum teleportation
and ``one way quantum computing'' would not work. Surprisingly, we
found that, in fact, von Neumann's description of measurements via
refinement implies (under natural assumptions) L\"uders projection
postulate. It seems that this important observation was missed
during last 70 years. This result closed the problem of the proper
use of the projection postulate in quantum information theory. One
can proceed with L\"uders postulate (as people in quantum
information really do).}

\section{Introduction}

Although the QI project approached the stage of technological (at
least experimental)  realizations, research on foundational
problems related to quantum information processing\footnote{See,
e.g., recent book of G. Jaeger \cite{Jaeger} and paper of M.
Asano, M. Ohya, and Y. Tanaka \cite{Ohya}.} did not become less
important. Moreover, many problems in foundations of QM which were
considered as of pure theoretical (or even philosophical) value
nowadays play an important role in (expensive) technological
projects. Thus such problems could not be simply ignored.
Development of QI also induces new approaches which foundational
basis should be carefully analyzed. Among such novel approaches I
would like to mention quantum teleportation and ``one way quantum
computing'', see, e.g., \cite{oneway}--\cite{oneway2}  -- an
exciting alternative to the conventional scheme of quantum
computing.

In a recent series of papers \cite{KH1}--\cite{KH4} the author
paid attention on crucial difference of consequences of von
Neumann \cite{VN} and L\"uders \cite{LUD} projection postulates
for QI, staring with EPR-argument \cite{EPR}. These postulates
coincide for observables with {\it nondegenerate spectra}, but
they differ in the case of {\it degenerate spectra.} We remark
that the latter case is the most important for quantum information
theory, since measurement on one of systems in a pair of entangled
systems is represented by an operator with degenerate spectrum.

While L\"uders \cite{LUD} projection postulate is fine for QI, the
appeal to von Neumann postulate induces serious problems \cite{}.
In the first case measurement on a subsystem produces a pure state
for another subsystem and it is good for quantum teleportation and
computing. However, in the second case even starting with a pure
state for a composite system, one obtains in general a statistical
mixture. Moreover, by von Neumann the formalism of QM does not
predict the post measurement state in the case of degenerate
spectrum. Thus even mentioned statistical mixture is unknown. In
\cite{VN} it was emphasized that measurementd of observables
represented by operators with degenerate spectra are {\it
ambiguous.} This problem can be solved (due to von Neumann) only
via {\it refinement measurements.} One should find an observable,
say $B,$ represented by an operator $\widehat{B}$ with
nondegenerate spectrum which commutes with the original operator
$\widehat{A}$ with degenerate spectrum. Then results of
$A$-measurement are obtained as $A=f(B),$  where $f$ is the
function coupling the operators: $\widehat{A}= f(\widehat{B}).$
Since $B$ can be chosen in various ways, one can select various
measurement procedures for $A$-measurement. It is crucial for
foundations of QI that for composite systems refinement of
measurement on one of subsystems can be approached only via
combined measurement on both subsystems. If it is really the case
and von Neumann was right, then foundations of QI should be
carefully reconsidered, since a number of important procedures in
QI processing is based on L\"uders postulate. First of all we
mention quantum teleportation. It were impossible to teleport an
unknown quantum state in von Neumann's framework, see \cite{VN}.
Alice evidently uses L\"uders postulate to be sure that her
measurement would produce the corresponding pure state for Bob
(then Bob needs only to perform a local unitary evolution to get
the proper state).

The situation in quantum computing is not completely clear. It
seems that the post-measurement state does not play any role in
the conventional scheme of quantum computation: unitary evolution
and, finally, measurement of a proper observable. It seems that
only probabilities of results are important. Probabilities are
calculated in the same way both in L\"uders and von Neumann's
approach. The situation is completely different in the case of so
called ``one way quantum computing'', see, e.g.,
\cite{oneway}--\cite{oneway2}. This scheme (based on measurements,
instead of unitary evolution) depends crucially on the possibility
to use L\"uders postulate. It would not work if von Neumann was
right and L\"uders was wrong.

To my surprise, recently I found that, in fact, von Neumann's
description of measurements via refinement\footnote{By using an
observable represented by an operator with nondegenerate spectrum
commuting with operator with degenerate spectrum representing the
original observable.} implies (under natural assumptions) L\"uders
projection postulate. It seems that this important observation was
missed during last 70 years. This result closed the problem of the
proper use of the projection postulate in quantum information
theory. One can proceed with L\"uders postulate (as people in
quantum information really do).

\section{Von Neumann's and L\"uders' postulates for pure states}

\subsection{Nondegenerate (discrete)  spectrum}

Everywhere below $H$ denotes  complex Hilbert space. Let $\psi \in
H$ be a pure state, i.e., $\Vert \psi \Vert^2=1.$ We remark that
any pure state induces density operator:
$$
\widehat{\rho}_{\psi}= \psi \otimes \psi= \widehat{P}_{\psi}
$$
where $\widehat{P}_{\psi}$ denotes the orthogonal projector on the
vector  $\psi.$ This operator describes an ensemble of identically
prepared systems each of them in the same state $\psi.$

 For an observable $A$  represented by the operator
 $\widehat{A}$ with nondegenerate
 spectrum von Neumann's and L\"uders projection postulates
 coincide. For simplicity we restrict our considerations to
 operators with {\it purely discrete spectra.} In
this case  spectrum consists of eigenvalues $\alpha_k$ of
$\widehat{A}: \widehat{A} e_k = \alpha_k e_k.$ Nondegeneracy of
spectrum means that {\it subspaces consisting of eigenvectors
corresponding to different eigenvalues are  one dimensional.}
 \medskip

{\bf PP:} {\it   Let $\; A\; $ be an observable described by the
self-adjoint operator $\widehat{A}$ having purely discrete
nondegenerate spectrum. Measurement of observable $\; A\; $ on a
system  in the (pure) quantum state $\psi$  producing the result
$A=\alpha_k$ induces transition from the state $\psi$ into the
corresponding eigenvector $e_k$ of the operator $\widehat{A}.$}

\medskip

If we select only systems with the fixed measurement result
$A=\alpha_k,$ then we obtain an ensemble described by the density
operator $\widehat{q}_k= e_k \otimes e_k.$ Any system in this
ensemble is in the same state $e_k.$  If we do not perform
selections, we obtain obtain an ensemble described by the density
operator
$$
\widehat{q}_\psi  =  \sum_{k} \vert \langle \psi, e_{k} \rangle
\vert^2 \widehat{P}_{e_{k}}=  \sum_{k} \langle \widehat{\rho}_\psi
e_{k}, e_{k} \rangle \widehat{P}_{e_{k}} =\sum_{k}
\widehat{P}_{e_{k}} \widehat{\rho}_\psi \widehat{P}_{e_{k}}.
$$
where $\widehat{P}_{e_{k}}$ is projector on the eigenvector
$e_{k}.$

\subsection{Degenerate (discrete)  spectrum: L\"uders viewpoint}

L\"uders generalized this postulate to the case of operators
having degenerate spectra. Let us consider spectral decomposition
for a self-adjoint operator $\widehat{A}$ having purely discrete
spectrum:
$$
\widehat{A}=\sum_i \alpha_i \widehat{P}_i,
$$
where $\alpha_i \in {\bf R}$ are different eigenvalues of
$\widehat{A}$ (so $\alpha_i \not=\alpha_j)$ and $\widehat{P}_i,
i=1,2,...,$ is projector onto subspace $H_i$ of eigenvectors
corresponding to $\alpha_i.$

By L\"uders' postulate after measurement of an observable $A$
represented by the operator $\widehat{A}$ that gives the result
$\alpha_i$ the initial pure state $\psi$ is {\it transformed again
into a pure state,} namely,
$$
\psi_i = \frac{\widehat{P}_i\psi}{\Vert \widehat{P}_i\psi \Vert}.
$$
Thus for corresponding density operator we have
$$
\widehat{Q}_i=  \psi_i \otimes \psi_i = \frac{\widehat{P}_i \psi
\otimes \widehat{P}_i \psi}{\Vert \widehat{P}_i\psi\Vert^2} =
\frac{\widehat{P}_i \widehat{\rho}_{\psi} \widehat{P}_i}{\Vert
\widehat{P}_i\psi\Vert^2}.
$$
If one does not make selections corresponding to values $\alpha_i$
the final post-measurement state is given by
\begin{equation}
\label{LVNzz} \widehat{q}_\psi = \sum_i p_i \widehat{Q}_i, \; \;
p_i= \Vert \widehat{P}_i\psi\Vert^2,
\end{equation}or simply
\begin{equation} \label{LVN3} \widehat{q}_\psi = \sum_i \widehat{q}_i, \; \widehat{q}_i= \widehat{P}_i \rho_\psi \widehat{P}_i.
\end{equation}
This is the statistical mixture of pure states $\psi_i.$  Thus by
{\it L\"uders there is no essential difference between
measurements of observables with degenerate and nondegenerate
spectra.}

\subsection{Degenerate (discrete)  spectrum: von Neumann's viewpoint}

Von Neumann had the completely different viewpoint on the
post-measurement state \cite{VN}. Even for a pure state $\psi$ the
post-measurement state (for measurement with selection with
respect to a fixed result $A=\alpha_k)$  will not be a pure state
again. If $\widehat{A}$ has degenerate (discrete) spectrum, then
according to von Neumann \cite{VN}

\medskip

{\it  A measurement of an observable $A$ giving the value
$A=\alpha_i$ does not induce projection of $\psi$ on the subspace
$H_i.$}

\medskip

 The result will not be the fixed pure state, in particular, not L\"uders' state
$\psi_i.$ Moreover, the post-measurement state, say
$\widehat{g}_\psi,$ is not determined by the formalism of QM! Only
a subsequent measurement of an observable $D$ such that $A=f(D)$
and $\widehat{D}$ is an operator  with nondegenerate spectrum
(``refinement measurement'') will determine the final state.

Following von Neumann, we choose in each $H_i$ an orthonormal
basis $\{e_{in}\}.$ Let us take sequence of real numbers
$\{\gamma_{in}\}$ such that all numbers are distinct. We define
the corresponding self-adjoint operator $\widehat{D}$ having
eigenvectors $\{e_{in}\}$ and eigenvalues $\{\gamma_{in}\}:$
$$
\widehat{D}= \sum_i \sum_n \gamma_{in} \widehat{P}_{e_{in}}.
$$
A measurement of the observable $D$ represented by the operator
$\widehat{D}$ can be considered as measurement of the observable
$A,$ because $A=f(D),$ where $f$ is some function such that
$f(\gamma_{in})=\alpha_i.$ The $D$-measurement (without
post-measurement selection with respect to eigenvalues) produces
the statistical mixture
\begin{equation} \label{LVN3A} \widehat{O}_{D;\psi}= \sum_{i} \sum_{n} \vert
\langle \psi, e_{in} \rangle \vert^2 \widehat{P}_{e_{in}}.
\end{equation}
By selection for the value $\alpha_i$ of $A$ (its measurements
realized via measurements of a refinement observable $D)$ we
obtain the statistical mixture described by normalization of the
operator
 \begin{equation} \label{LVkk}
\widehat{O}_{i,D;\psi} =  \sum_{n} \vert \langle \psi, e_{in}
\rangle \vert^2 \widehat{P}_{e_{in}}.
\end{equation}

Von Neumann  emphasized that the {\it mathematical formalism of QM
could not describe the post-measurement state for measurements
(without refinement)  of degenerate observables.} He did not
discuss directly properties of such a state, he described them
only indirectly via refinement measurements.\footnote{For him this
state was a kind of hidden variable. It might even be that he had
in mind that it ``does not exist at all'', i.e., it could not be
described by a density operator.} We would like to proceed by
considering this (``hidden'') state under assumption that it can
be described by a density operator, say $\widehat{g}_\psi.$ We
formalize a list of properties of this hidden (post-measurement)
state which can be extracted from von Neumann's considerations on
refinement measurements. Finally, we prove, see Theorem 1, that
$\widehat{g}_\psi$ should coincide with the post-measurement state
postulated by L\"uders, (\ref{LVN3}).

Consider the $A$-measurement without refinement. By von Neumann,
for each quantum system $s$ in the initial pure state $\psi,$ the
$A$-measurement with the $\alpha_i$-selection transforms the
$\psi$ in one of states $\phi=\phi(s)$ belonging to the
eigensubspace $H_i.$ Unlike L\"uders' approach, it implies that,
instead of one fixed state, namely, $ \psi_i \in H_i,$ such an
experiment produces a probability distribution of states on the
unit sphere of the subspace $H_i.$ We postulate

\medskip

{\bf DO} For any value $\alpha_i$ such that $\widehat{P}_i\psi
\not=0,$ the post-measurement probability distribution on $H_i$
can be described by a density operator, say $\widehat{\Gamma}_i.$

\medskip

Here $\widehat{\Gamma}_i: H_i \to H_i$ is such that
$\widehat{\Gamma}_i \geq 0$ and $\rm{Tr} \widehat{\Gamma}_i=1.$
Consider now the corresponding density operator $\widehat{G}_i$ in
$H.$ Its restriction on $H_i$ coincides with $\widehat{\Gamma}_i.$
In particular this implies its property:
\begin{equation} \label{LVe}
\widehat{G}_i(H_i)  \subset H_i.
\end{equation}
We remark that $\widehat{G}_i$ is determined by $\psi,$ so
$\widehat{G}_i\equiv \widehat{G}_{i; \psi}.$

We would like to present the list of other properties of
$\widehat{G}_i$ induced by von Neumann's considerations on
refinement. Since,  for each refinement measurement $D,$ the
operators $\widehat{A}$ and $\widehat{D}$ commute, the measurement
of $A$  with refinement can be performed in two ways. First  we
perform the $D$-measurement and then we get $A$ as $A=f(D).$
However, we also can first perform the $A$-measurement, obtain the
post-measurement state described by the density operator
$\widehat{G}_i,$  then measure $D$ and, finally, we again find
$A=f(D).$

Take an arbitrary  $\phi \in H_i$ and consider a refinement
measurement $D$ such that $\phi$ is an eigenvector of
$\widehat{D}.$ Thus $\widehat{D} \phi = \gamma_\phi \phi.$ Then
 for the cases -- [direct measurement of $D$] and [first $A$ and then $D]$ -- we  get
probabilities which are coupled in a simple way. In the first case
(by Born's rule) \begin{equation} \label{LVkk1} {\bf
P}(D=\gamma_\phi \vert \widehat{\rho}_\psi)= \vert <\psi,
\phi>\vert^2.
\end{equation}
In the second case, after the $A$-measurement, we obtain the state
$\widehat{G}_i$ with probability $${\bf P}(A=\alpha_i \vert
\widehat{\rho}_\psi)= \Vert \widehat{P}_i \psi \Vert^2.$$
Performing the $D$-measurement for the state $\widehat{G}_i$ we
get the value $\gamma_\phi$ with probability: \begin{equation}
\label{LVkk2} {\bf P}(D=\gamma_\phi \vert \widehat{G}_i)=
\rm{Tr}\; \widehat{G}_i \widehat{P}_\phi.
\end{equation}
By (classical) Bayes' rule
\begin{equation} \label{LVkk3} {\bf
P}(D=\gamma_\phi \vert \widehat{\rho}_\psi)= {\bf P}(A=\alpha_i
\vert \widehat{\rho}_\psi) {\bf P}(D=\gamma_\phi \vert
\widehat{G}_i).
\end{equation}
Finally, we obtain
\begin{equation} \label{LVkk4}
{\bf P}(D=\gamma_\phi \vert \widehat{G}_i)= \rm{Tr}
\;\widehat{G}_i \widehat{P}_\phi =\frac{\vert <\psi,
\phi>\vert^2}{\Vert \widehat{P}_i \psi \Vert^2}.
\end{equation}
Thus
\begin{equation} \label{LVkk5}
\rm{Tr} \widehat{G}_i \widehat{P}_\phi =\frac{\vert <\psi,
\phi>\vert^2}{\Vert \widehat{P}_i \psi \Vert^2}.
\end{equation}
This is one of the basic features of the post-measurement state
$\widehat{G}_i$ (for the $A$-measurement with the
$\alpha_i$-selection, but without any refinement). Another basic
equality we obtain in the following way. Take an arbitrary
$\phi^\prime \in H_i^{\perp},$ and consider a  measurement of the
observable described by the orthogonal projector
$\widehat{P}_{\phi^\prime}$ under the state $\widehat{G}_i.$ Since
the later describes a probability distribution concentrated on
$H_i,$ we have:
\begin{equation} \label{LVkki}
{\bf P}(P_{\phi^\prime}= 1 \vert \widehat{G}_i)= 0.
\end{equation}
Thus \begin{equation} \label{LVkkii} \rm{Tr} ;\ \widehat{G}_i
\widehat{P}_{\phi^\prime} = 0.
\end{equation}
This is the second basic feature of the post-measurement state.
Our aim is to show that (\ref{LVkk5}) and (\ref{LVkkii}) imply
that, in fact,
\begin{equation} \label{LVkkj}
\widehat{G}_i= \widehat{P}_i \widehat{\rho}_\psi
\widehat{P}_i/\Vert \widehat{P}_i \psi\Vert^2 \equiv \widehat{P}_i
\psi \otimes \widehat{P}_i \psi/\Vert \widehat{P}_i \psi\Vert^2 ,
\end{equation}
i.e., to derive L\"uders postulate which is a theorem in our
approach.

{\bf Lemma.} {\it The post-measurement density operator
$\widehat{G}_i$ maps $H$ into $H_i.$}

{\bf Proof.} By (\ref{LVe}) it is sufficient to show that
$\widehat{G}_i(H_i^{\perp}) \subset H_i.$  By (\ref{LVkkii}) we
obtain
\begin{equation} \label{LVll}
< \widehat{G}_i \phi^\prime,\phi^\prime>=0
\end{equation}
for any $\phi^\prime \in H_i^{\perp}.$ This immediately implies
that $< \widehat{G}_i \phi_1^\prime,\phi_2^\prime>=0$ for any pair
of vectors from $H_i^{\perp}.$ The latter implies that
$\widehat{G}_i \phi^\prime \in H_i$ for any $\phi^\prime \in
H_i^{\perp}.$

\medskip

Consider now the $A$-measurement without refinement and selection.
The post-measurement state $\widehat{g}_\psi$ can be represented
as
\begin{equation} \label{LVmnn}
\widehat{g}_\psi = \sum_m p_m \widehat{G}_m, \; \; p_= \Vert
\widehat{P}_m \psi\Vert^2,
\end{equation}

{\bf Proposition 1.} {\it For any pure state $\psi$ and
self-adjoint operator $\widehat{A}$ (with purely discrete
degenerate) spectrum the post-measurement state (in the absence of
refinement measurement) can  be represented as
\begin{equation} \label{LVmmj}
\widehat{g}_\psi = \sum_m \widehat{g}_m,
\end{equation}
where $\widehat{g}_m: H \to H_m, \widehat{g}_m \geq 0,$ and, for
any $\phi \in H_m,$}
\begin{equation} \label{LVmk}
<\widehat{g}_m \phi, \phi>= \vert <\psi, \phi>\vert^2.
\end{equation}

\section{Derivation of Luders' postulate from von Neumann's
postulate}

{\bf Theorem.} {\it Let $ \widehat{g}\equiv \widehat{g}_\psi$ be a
density operator described by Proposition 1. Then}
\begin{equation}
\label{**1} \widehat{g}_m=\widehat{P}_m \psi \otimes \widehat{P}_m
\psi.
\end{equation}

{\bf Proof.} Let $\{e_{mk}\}$ be an orthonormal basis in $H_m$ and
let $u \in H.$ We represent it as $u=u_m + u_m^{\perp},$ where
$u_m \in H_m$ and $u_m^{\perp} \in H_m^{\perp}.$ Then
$$
<\widehat{g}_m u, u>= <\widehat{g}_m u_m, u_m> +<\widehat{g}_m
u_m, u_m^{\perp}> + <\widehat{g}_m u_m^{\perp}, u_m > +
<\widehat{g}_m u_m^{\perp}, u_m^{\perp}>.
$$
The second  and last terms equals to zero, since $\widehat{g}_m: H
\mapsto H_m.$ To show that the third term also equals to zero, we
should use self-adjointness of $\widehat{g}_m.$ Thus
$$
<\widehat{g}_m u, u> = \sum_{k, k^\prime} <u, e_{ku}>
<e_{mk^\prime}, u> <\widehat{g}_m e_{ku}, e_{mk^\prime}>.
$$
For each $e_{mn},$ we have $<\widehat{g}_m e_{mn}, e_{mn}>=|<\psi,
e_{mn}>|^2.$ Thus the diagonal elements of the matrix of operator
$\widehat{g}_m$ coincide with diagonal elements of operator
$\widehat{P}_m \psi \otimes \widehat{P}_m \psi.$ Take now another
basis in $H_m$ which is constructed in the following way. We fix
two indexes, say $n$ and $j$, and choose two new basis vectors:
$$
f_{mn}= (e_{mn} + e_{mj})/\sqrt{2}, \; f_{mj}= (e_{mn} -
e_{mj})/\sqrt{2}.
$$
Then we have $<\widehat{g}_m f_{mn}, f_{mn}>=|<\psi, f_{mn}>|^2,$
or
$$<\widehat{g}_m e_{mn}, e_{mn}> + <\widehat{g}_m e_{mj}, e_{mj}> + <\widehat{g}_m e_{mn},
e_{mj}> + <\widehat{g}_m e_{mj}, e_{mn}>
$$
$$
= |<\psi, e_{mn}>|^2 + |<\psi, e_{mj}>|^2 + <\psi, e_{mn}><e_{mj},
\psi> + <\psi, e_{mj}><e_{mn}, \psi>.
$$ Thus $$ <\widehat{g}_m e_{mn}, e_{mj}>
+ \overline{<\widehat{g}_m e_{mn}, e_{mj}>} $$ $$ = <\psi,
e_{mn}><e_{mj}, \psi> + \overline{<\psi, e_{mn}><e_{mj}, \psi>}.
$$

Thus we proved that Re $[<\widehat{g}_m e_{mn}, e_{mj}>]=$ Re
$[<\psi, e_{mn}><e_{mj}, \psi>].$ Let us now choose two new basis
vectors
$$
\bar f_{mn}= (e_{mn} + i e_{mj})/\sqrt{2}, \; , \bar f_{mj}= (e_{mn} + i e_{mj})/\sqrt{2}.
$$
Then we have:
$$<\widehat{g}_m \bar f_{mn}, \bar f_{mn}>=
<\widehat{g}_m e_{mn}, e_{mn}> + <\widehat{g}_m e_{mj}, e_{mj}> $$
$$
 + i <\widehat{g}_m e_{mj},
e_{mn}> - i <\widehat{g}_m e_{mn}, e_{mj}> =
$$
$$
|<\psi, e_{mn}>|^2 +  <\psi, e_{mj}>|^2 + i <\psi, e_{mn}><e_{mj},
\psi> - i <\psi, e_{mj}> <e_{mn}, \psi>.
$$
Thus:
$$
<q_{m}e_{mj}, e_{mn}> - <\widehat{g}_m e_{mn}, e_{mj}> = <\psi,
e_{mn}><e_{mj}, \psi>-<\psi, e_{mj}><e_{mn}, \psi>.
$$

Thus $<q_{m} e_{mn}, e_{mj}>= <\psi, e_{mj}><e_{mn}, \psi>.$ We
obtained the following representation for the quadratic form of
the operator $\widehat{g}_m$
$$
<\widehat{g}_m u, u>= \sum_{k, k^\prime} <u, e_{mk}>
<e_{mk^\prime}, u> <\psi, e_{mk^\prime}> <e_{mk}, \psi>= |<\psi,
u>|^2.$$ Hence $\widehat{g}_m=\widehat{P}_m \psi \otimes
\widehat{P}_m^\psi.$

\medskip

{\bf Conclusion:} {\it The general scheme of measurement of
observables with degenerate spectra provided by von Neumann
\cite{VN} implies, in fact,  the L\"uders projection postulate.
This postulate is a theorem (missed for 70 years) in von Neumann's
framework. Thus (in the canonical formalism of QM)  the
post-measurement state is always a pure state. This supports
existing schemes of quantum teleportation and computing.}

\end{document}